\documentclass[aps,epsf,preprint]{revtex4}

\def\1{{\bf 1}}
\def\[{\left[}
\def\]{\right]}
\def\be{\begin{eqnarray}}
\def\ee{\end{eqnarray}}
\def\nn{\nonumber}
\def\({\left(}
\def\){\right)}
\def\bk#1{\langle#1\rangle}
\def\eq#1{(\ref{#1})}

\def\o{\omega}

\def\f{\phi}

\def\G{{\cal G}}
\def\l{\lambda}
\def\m{\mu}
\def\Tr{{\rm Tr}}
\def\r2{\sqrt{2}}
\def\n{\noindent}
\def\C{{\cal C}}

\begin{document}

\title{A Higgs Test of Horizontal Symmetry}
\author{C.S. Lam}
\address{Department of Physics, McGill University\\
 Montreal, Q.C., Canada H3A 2T8\\
and\\
Department of Physics and Astronomy, University of British Columbia,  Vancouver, BC, Canada V6T 1Z1 \\
Email: Lam@physics.mcgill.ca}

\begin{abstract}
Identical interactions found in the three families of quarks and leptons suggest the presence of a
horizontal symmetry. We discuss how  such a symmetry can be detected by measuring the
decay rates of Higgs into fermion pairs, and the Higgs production cross section. Depending on the
details, there is a chance  that
the decay widths to the bottom-pair and the tau-pair may be down by more than  a factor of 3  compared to the usual values,
and the fusion production cross section of the Higgs also altered. Whatever the outcome, such a test also serves
to constraint horizontal symmetry models.
\end{abstract}
\narrowtext
\maketitle

Ever since the famous utterance of I.~I. Rabi, demanding to know who ordered the muon when it was discovered
in 1936, the presence
of several families of quarks and  leptons with identical interactions has remained an enigma. The triplication of families
naturally suggests a symmetry among them, usually known as a horizontal symmetry, yet a direct experimental confirmation of this hypothesis
is difficult because the symmetry 
is expected to be broken in order  to account for
the fermion mass differences and  mixing. If we assume the breaking to be spontaneous and triggered by scalar condensates, then the  presence of 
 scalar fields could serve as an experimental signature for the symmetry, but again most of these scalar fields
might turn out to be too heavy to be detectable in the foreseeable future.  Nevertheless, the presence of their condensates
 may still serve as a tool for us to infer on the existence of a horizontal symmetry.

The regularity of neutrino mixing, the so-called `tri-bimaximal mixing' \cite{HPS}, has inspired many explanations from horizontal-symmetry models.
 Some of these models also deal with fermion masses and quark mixing. A skeptic 
may however not take the success of these models to be evidence for the presence of a horizontal symmetry, for at least two reasons.
First, these models all contain a number of adjustable parameters. With enough parameters, one can always explain all the
experimental data even without a horizontal symmetry, so it is not clear to what extent the success of these models actually reflects the presence of
a symmetry. Secondly, if such a symmetry is indeed present in nature, it should be unique. Yet many symmetry groups, groups including 
$S_3, A_4, T', S_4, PSL(2,7), \Delta(2n^2), \Delta(6n^2), D_n, Q_n, Z_n, SO(3), SU(3)$, have been invoked to explain the data \cite{GPS}. 
The success over such a large
range of groups might cause concern on the real presence of a unique symmetry.

We propose in this note an experimental test for the presence of horizontal symmetry,  independent of these
model details, once the Higgs particle currently being intensely searched for is found. It consists of measuring
its decay widths to various fermion pairs, and its production cross section accurately.

As is well known, precision electroweak data suggest that one of these scalar particles, the one that gives rise to the $W$ and $Z$ masses, may be discovered in the 
near future \cite{W1}. We shall refer to this scalar particle as the Standard-Model (SM) Higgs, and denote it by $H$. 
In the presence of a horizontal symmetry,
the fermions are expected to carry horizontal quantum numbers to distinguish those of one family from another, but the gauge bosons, which are
common to all three families and couple equally to all the fermion pairs,  should be  horizontal singlets. What about the SM Higgs
that gives rise to the masses of these gauge bosons? The simplest assumption is that it is also a horizontal singlet, an assumption which we shall make in the rest
of the article for the sake of concreteness. However, even if the SM Higgs is not a singlet but a member of a multiplet, it still does not detract from the main point of
this note, which is that the coupling between the SM Higgs and the fermion pairs is {\it not} proportional to the mass of the fermion if horizontal symmetry exists,
so the decay rates of the SM Higgs to fermion pairs, partially relied on for the detection of the SM Higgs, are not equal to the usual SM estimates.

In the SM, fermion mass terms are of the form
\be
&&f_{ij}\overline E_i\ell_jH^\dagger +g_{ij}\overline N_i\ell_jH-{1\over 2}M^N_{ij}N_iN_j\nn\\
&+&f'_{ij}\overline D_iq_jH^\dagger +g'_{ij}\overline U_iq_jH +h.c.,\label{mass}\ee
where $\ell_i=\pmatrix{e_{Li}\cr \nu_{L_i}\cr}$ denote the three families ($i=1,2,3$) of left-handed leptons, and $q_i=\pmatrix{u_{Li}\cr d_{Li}\cr}$
denote the three families of left-handed quarks. The right-handed leptons are denoted by $E_i, N_i$, and the right-handed quarks by $U_i, D_i$.
For simplicity, I assume in \eq{mass} the existence of $N_i$ and a Majorana mass term $M^N_{ij}$, but this is not crucial to the test, especially
because neutrinos are probably unobservable anyway.  
$H$ is the SM Higgs, and $f_{ij}, g_{ij}, f'_{ij}, g'_{ij}$ are the Yukawa coupling constants which specify the various Dirac mass matrices. 

From now on I will concentrate on the first term of \eq{mass}; the discussion and the conclusion  reached for the other terms are essentially identical.
In the presence of a spontaneously-broken horizontal symmetry $\G$, suppose $\overline E$ belongs to a $\G$-representation $I$, and $\ell$ to a $\G$-representation $J$. In order to keep the mass term $\G$-invariant before symmetry breaking, 
$f_{ij}$ should be of the form 
\be
f_{ij}=\sum_A y_A(Ii,Jj|Aa)\bk{\f^A_a},\label{hs}\ee
where $\f^A_a$ is a scalar object
 in the irreducible representation $A$, $(Ii,Jj|Aa)$ is the Clebsch-Gordan (CG) coefficient coupling $\overline E^I_i\ell^J_j$ to $\f^{A*}_a$
to render $(Ii,Jj|Aa)\overline E^I_i\ell^J_j\f^A_a$  invariant under $\G$. The sum is over all irreducible representations $A$, and a
summation over the repeated indices $i,j,a$ is understood. 
Before symmetry breaking takes place, the first term of \eq{mass} is then replaced by a $\G$-invariant expression
\be
\sum_Ay_A(Ii,Jj|Aa)\overline E^I_i\ell^J_j\f^A_aH^\dagger,\label{invariant}\ee
where $y_A$ are adjustable coupling constants. There must be at least three of them to enable the three charged-lepton masses
to be fitted. It is therefore clear that
beyond the SM Higgs, there must be additional scalars present with non-trivial horizontal quantum numbers. Their condensates also
contribute to the mass matrices,  not just that of the SM Higgs as in the Standard Model. It is this difference that allows us to
test the presence of a horizontal symmetry.

In order for the mass term to be renormalizable, the combination $\f^A_aH^\dagger$ in \eq{invariant} must appear
as a single scalar field $H^{A\dagger}_a$, transforming as an isodoublet and a $\G$-Aplet. 
$A=\1$ will be used to designate
 the singlet,
hence the SM Higgs is $H^\dagger=H^{\1 \dagger}$. The charged-lepton mass matrix is 
\be
M_{ij}=\sum_Ay_A(Ii, Jj|Aa)\bk{H^{A\dagger}_a}:=\sum_AM^A_{ij},\label{mm}\ee
where $\bk{H^{A\dagger}_a}$ denotes the vacuum expectation value of the electrically neutral component, with possible isospin CG coefficients absorbed in it. 

Let us remind ourselves of some basic properties of the CG coefficients $(Ii,Jj|Aa)$ when $I,J,A$ are irreducible. The coefficients satisfy the unitarity relation
\be
(Ii,Jj|Aa)^*(Ii,Jj|A'a')=\delta_{AA'}\delta_{aa'},\label{uni}\ee
 and possess the group-transformation property
\be D^I_{ii'}(g)D^J_{jj'}(g)(Ii',Jj'|Aa)=D^{A}_{aa'}(g)(Ii,Jj|Aa')\label{gp}\ee
for every $g\in \G$. The irreducible representations $D^I, D^J, D^A$ will be taken to be unitary.

It follows from \eq{mm} and \eq{uni} that
\be \Tr(M^\dagger M)=\sum_{A,a}|y_A\bk{H^{A\dagger}_a}|^2\equiv |y_\1\bk{H^{\1 \dagger}}|^2(1+R)=m_e^2+m_\mu^2+m_\tau^2\equiv 3\overline{m^2},\label{trace}\ee
where $R$ is the ratio of the horizontal non-singlet to singlet contributions. It measures the reduction of the total decay width of Higgs
to all fermion pairs, as we shall see. It will also become clear shortly that $R\not=0$, or else $m_\tau=m_\mu=m_e$, which is false. The size of $R$ is somehow
related to the mass differences, so it cannot be too small,
but its precise value is model dependent, though it can be measured from the decay rates.
For example, in models where the Higgs condensates can be chosen to make $M_{ij}$ diagonal, and there are many such models, then $R$ can be easily computed from
the left hand side of \eq{trace} to be
\be 1+R={3(m_e^2+m_\m^2+m_\tau^2)\over (m_e+m_\m+m_\tau)^2},\label{R}\ee
if the charged leptons belong to an irreducible triplet.

For $A=\1, D^A=\1$, \eq{gp} can be written as
\be D^I_{ii'}(g)(Ii',Jj|\1 1)=(Ii,Jj'|\1 1)D^{J*}_{j'j}(g).\label{a1}\ee
Together with \eq{uni}, this formula shows that  the irreducible representation $J^*$ is either equivalent to $I$, or $(Ii,Jj|\1 1)=0$ so
there is no Higgs decay to the pair. In the former case, we can put $J^*=I$, then \eq{a1}  becomes
$
D^I_{ii'}(g)(Ii',I^* j\1 1)=(Ii,I^*j'|\1 1)D^I_{j'j}(g).$
By Schur's lemma, one concludes that 
the CG coefficient is proportional to $\delta_{ij}$. The normalization in \eq{uni} then tells us that
$(Ii,I^*j|\1 1)=\delta_{ij}/\sqrt{n_I}$, where $n_I$
is the multiplicity of the irreducible multiplet $I$, which is 3 in the present case, but the formula is true whatever the multiplicity of $I$ is. 

Thus, assuming the left-handed leptons $\ell$ to form an irreducible triplet, we conclude that the SM Higgs will decay 
into a charged-lepton pair only when the right-handed charged leptons $E$ also belong to
the same irreducible triplet. In that case, the Higgs-lepton-pair coupling is equal to $y_\1/\sqrt{3}$ for every pair, and the partial width for Higgs decaying
into each pair is proportional to $|y_\1|^2/3=\overline{m^2}/(1+R)|\bk{H^{\dagger}}|^2$, according to \eq{trace}. The corresponding
prediction in the SM is $m_f^2/|\bk{H^{\dagger}}|^2$, where $m_f$ is the fermion mass of the pair in question. These two differ by a factor 
$F=\overline{m^2}/m_f^2(1+R)$.

The same discussion  applies to the up-type and down-type quarks if they are irreducible horizontal triplets. The factor $R$ however may
be different for these other cases.

Since $m_\tau^2/\overline{m^2}\simeq 3$, the tau-pair decay rate is down from the SM rate by  a
factor  $\simeq 3(1+R)$.  In particular, if \eq{R} holds, then the factor is close to 9.
What is lost in the tau-pair rate is 
partially compensated by the increase in
mu-pair and e-pair rates, but the total decay width into these three pairs is still a factor $(1+R)$ down from the SM width.

The same argument applies to the bottom quark if it is a member of an irreducible horizontal triplet: the SM Higgs to b-pair decay rate is down by a factor 
$\simeq 3(1+R)$ compared to the SM rate \cite{DEC}. This reduction would make the experimental detection of Higgs more difficult \cite{ATLAS}. 

If such a reduction of Higgs to tau-pair or Higgs to b-pair decay rate
is observed, by comparing the measured rate against the computed rate available in the literature \cite{DEC},
then it is a pretty strong signal that a horizontal symmetry exists, and the corresponding
triplet is irreducible. We can then rule out horizontal groups without a 3-dimensional irreducible representation,
groups such as $S_3, D_n$, and all the abelian groups. If a reduction is not seen in tau-pair decay, then we can rule out all models assuming either $E$ or $\ell$ to an irreducible
triplet, which are most of the models with a symmetry group possessing a 3-dimensional irreducible representation. 
By the same argument, it also rules out models in which tau is a member of an irreducible doublet. 
Similarly, if a reduction is not seen in b-pair decay, then we can rule out all models assuming either $D$ or $q$ to an irreducible triplet or doublet. 
 Models assigning $E$ and $\ell$ to non-conjugate
irreducible multiplets ($I\not=J^*$) should see no coupling to the SM Higgs and no decay into charged-lepton pairs.
Models assigning $D$ and $q$ to non-conjugate irreducible multiplets would predict no decay of the SM Higgs
into b-pairs. We see therefore that whatever is the outcome of the measurement, it serves to rule out many of the existing horizontal symmetry models.

This change in Higgs coupling from the SM value also affects the Higgs 
production cross section, most of which according to the current estimate comes from gluon-gluon fusion \cite{PROD}.
Instead of receiving fusion contribution mostly from the production and subsequent annihilation of the top pair,
now charm pair and up pair also contribute equally when we ignore their kinematical differences. What cross section is lost from the top pair 
fusion is partly compensated
by the charm and up pairs, but the total fusion production cross section is still down by a factor $1+R$.
Similar discussions can be applied
to fusion production via the bottom pair. With light-quark contributions increased, Higgs production via direct annihilation of quarks and
anti-quarks in the partons also increases. It requires a detailed model calculation to see how the total Higgs production cross section 
changes, but whatever it is, it is probably not the same as before. By comparing such a calculation with an accurate experimental measurement of the 
cross section, there is also hope of seeing the presence or absence of horizontal symmetry in this way.

Only tree-order effects have been taken into account in these considerations. If experimental rates are indeed found to be 
different from the SM estimates, then higher-order corrections should be calculated to get the exact rates and cross sections.

 Similar discussion applies if a fermion belongs to an irreducible doublet. The SM Higgs decay rate to this fermion is 
then a factor $F'=\overline{{m'}^2}/m_f^2(1+R)$ of the SM rate, where $\overline{{m'}^2}$
is the average mass-squared of the doublet. Similarly, the gluon fusion cross section via this fermion pair also differs
from the SM cross section by the same factor.

If a fermion is a singlet, then both the decay rate and the fusion cross section is identical to those computed from the SM.

Since there are 6 leptons and 6 quarks, in principle by measuring the SM Higgs decay rate to each of these 12 pairs, one can determine whether the corresponding fermion
is a member of an irreducible triplet, doublet, or singlet. If the SM Higgs decay rates are observed for all of them,
then either no horizontal symmetry exists, or if it does, it is given by an abelian group.
In practice, other than the b-pair and perhaps the tau-pair, the measurement may be very difficult. 
Even so, these two decay channels alone already provide us with invaluable information regarding horizontal symmetry.

In conclusion, we have pointed out that if horizontal symmetry exists, then the decay rates of the SM Higgs to fermion pairs and the fusion production rate of the Higgs
are not the same as those given in the literature. A measure of these rates would place a great deal of restrictions on possible horizontal symmetry groups and horizontal symmetry models. We have worked out in detail the case when the SM Higgs is a horizontal singlet, but similar calculations can be carried out when it belongs to some other
multiplet by using the appropriate Clebsch-Gordan coefficients.


\begin{thebibliography}{9}
\bibitem{HPS} P.F. Harrison, D.H. Perkins, and W.G. Scott, Phys.~Lett. B458,  (1999) 79,
hep-ph/9904297; Phys.~Lett. B530, (2002) 167, hep-ph/0202074.
\bibitem{GPS} It is difficult to cite the more than 200 papers dealing with the various groups that can be found in
the archive http://arXiv.org. At the time
of writing, the number of papers dealing with $A_4$ already exceeds 120.
\bibitem{W1} See http://lepewwg.web.cern.ch/LEPEWWG/Welcome.html for a summary.
\bibitem{DEC} M. Carena and H.E. Haber, Prog. in Part.~Nucl.~Phys. 50 (2003) 152.
\bibitem{ATLAS} J.A. Valls Ferrer, J.~Phys. Conference Series 171 (2009) 012026.
\bibitem{PROD} For a summary of production  cross sections at Tevatron and LHC energies, see http://maltoni.home.cern.ch/maltoni/TeV4LHC/SM.html.

\end{thebibliography}
\end{document}